\documentclass{osa-article}
\journal{oe}

\usepackage{graphicx}
\usepackage{caption}
\usepackage{subcaption}
\usepackage{multibib}
\usepackage{float}
\usepackage{xcolor}
\usepackage{soul}
\usepackage[utf8]{inputenc}     
\protect

\protect
\usepackage{ulem}
\usepackage{braket}
\usepackage{multirow}
\begin{document}
\title{Polarimetric imaging of the human brain to determine the orientation and degree of alignment of nerve fiber bundles}

\author{Arushi Jain\authormark{1}, Leonie Ulrich\authormark{1}, Michael Jaeger\authormark{1}, Philippe Schucht\authormark{2},  Martin Frenz\authormark{1,*} and H. Günhan Akarcay\authormark{1}}	
\address{\authormark{1}Biomedical Photonics Department, Institute of Applied Physics, University of Bern,  Sidlerstrasse 5
	CH-3012 Bern, Switzerland\\
\authormark{2}Department of Neurosurgery, University hospital Bern, Freiburgstrasse 16
CH-3010 Bern, Switzerland\\
\email{\authormark{*}Corresponding author: martin.frenz@iap.unibe.ch}}\


\begin{abstract}
	More children and adults under the age of 40 die of brain tumor than from any other cancer. Brain surgery constitutes the first and decisive step for the treatment of such tumors. It is extremely crucial to achieve complete tumor resection during surgery, however, this is a highly challenging task, as it is very difficult to visually differentiate tumorous cells from the surrounding healthy white matter.
	The nerve fiber bundles constitutive of the white matter are organized in such a way that they exhibit a certain degree of structural anisotropy and birefringence. The birefringence exhibited by such aligned fibrous tissue is known to be extremely sensitive to small pathological alterations. Indeed, highly aligned anisotropic fibers exhibit higher birefringence than structures with weaker alignment and anisotropy, such as cancerous tissue. 
	
	In this study, we performed experiments on thick coronal slices of a healthy human brain to explore the possibility of (i) measuring, with a polarimetric microscope (employed in the backscattering geometry to facilitate non-invasive diagnostics), the birefringence exhibited by the white matter and (ii) relating the measured birefringence to the fiber orientation and the degree of alignment. This is done by analyzing the spatial distribution of the degree of polarization of the backscattered light and its variation with the polarization state of the probing beam. 
	
We demonstrate that polarimetry can be used to reliably distinguish between white and gray matter in the brain, which might help to intraoperatively delineate unstructured tumorous tissue and well organized healthy brain tissue. In addition, we show that our technique is able to sensitively reconstruct the local mean nerve fiber orientation in the brain, which can help to guide tumor resections by identifying vital nerve fiber trajectories thereby improving the outcome of the brain surgery.  	
	 \\
\end{abstract}

\section{Introduction}
\label{sec:section1}
Cancer is a global burden that leads to public health and economic problems and ranks as the first or second leading cause of premature death ~\cite{Atun2018}. Brain tumors, in particular the malignant ones, were responsible for  238'000 new cases in 2008 worldwide ~\cite{Ferlay2010, Louis2007}. The primary treatment for a brain tumor is surgery followed by adjuvant chemo or radio therapy.  Gross total resection, \textit{i.e.}, removal the bulk of cancer cells, is paramount for an improved prognosis. Hence, the accuracy of the estimation of the tumor's borders carries heavy consequences for the patients: on one hand, subtotal resection leads to increased risk of recurrence and shorter life expectancy~\cite{Violaris2012,Gotay2008,Stummer2008} and, on the other hand, if healthy tissue is removed as a result of gross resection, this might irrevocably damage the brain, leading to irreversible disabilities and a decreased quality of life which again might decrease life expectancy~\cite{McGirt2009}. The accurate resection of the malignant tumor is also a fundamental prerequisite for efficient adjuvant therapy~\cite{Lacroix2001,Sanai2008}. The prime difficulty resides in identifying clear borders to the tumorous cells to define adequate safety margins given the infiltrative nature of most brain tumor types  ~\cite{Claes2007, Belsuzarri2016}.

Current intraoperative modalities to delineate the tumor include visualization tools such as image guided neuronavigation, magnetic resonance imaging, computed tomography, or ultrasound ~\cite{Senft2011, Orringer2012, Prada2014, Charalampaki2015}, fluorescence guided techniques ~\cite{Martirosyan2011, Vorst2012, VanDam2011, Stepp2018}, and histopathology. Unfortunately, these modalities, having significantly improved tumor detection are still unable to reliably differentiate cancerous from healthy tissue  ~\cite{mohs2010,Stieglitz2013, Holland2000, Keereweer2013}.

 Polarimetric light imaging is a promising alternative to aforementioned techniques \cite{Schucht2020, RamellaRoman2020, Menzel2020a} . White matter in the mammalian brain primarily comprises of nerve fiber bundles ~\cite{Blumenfeld2018}
 that exhibit a certain degree of structural anisotropy and anisotropy in dielectric response~\cite{Tuchin2016}. These effects are coupled and manifest themselves via optical birefringence~\cite{Courtney2006}. Consequently, polarimetric light imaging has been used to reconstruct three-dimensional architecture of nerve fiber bundles and to map the nerve fibers orientations in and out of the plane of histological sections of a human brain ~\cite{Axer1999, Axer2000a, Axer2000, Axer2001, Axer2011a, Axer2011b, Dammers2012}.  Imaging the nerve fibers and their connectivity is also of great interest to construct the human connectome in order to unravel the architecture and connectivity of the nerve fiber bundles~\cite{Behrens2012}. 

Conversely, the polarimetric response of brain tissue can be used to detect disease-related derangement of the white matter's structural integrity. It has, for example, been shown that the white matter tracts in children (six months) affected by autism spectrum disorder show aberrant development~\cite{Wolff2013}. Similarly, the anisotropy is observed to be weaker in a cancerous or scar tissue than in health tissue ~\cite{Ghosh2010, Wood2009}, allowing to observe even small alterations via polarimetry ~\cite{Wallenburg2010,Aitken2012}. Hence, there is potential for polarimetric imaging to be used for non-invasive, non-contact, and real-time intra-operative cancer delineation.

As a first step towards this goal, we focus in this study on demonstrating that backscattering polarimetric microscopy is sensitive to detect structural differences in nerve fiber architecture in the human brain. The methodology follows the one presented in~\cite{Hornung2019,Jain2020}, where the spatial distribution of polarization of diffusely backscattered light is analyzed as a function of input polarization of a weakly focused probing beam. This allows us to assess the samples' structural anisotropy and thus give an indication of its degree of alignment. We proved that the degree of linear polarization can be used as a measure for determining the degree of alignment of fibrous materials. Although our final goal is to distinguish between tumorous tissue and healthy white matter, here, our aim is to test the methodology on non-tumorous formalin fixed (from now referred to as healthy) brain tissue. Firstly, we demonstrate the ability to distinguish between the highly aligned nerve fiber architecture in white matter and the low degree of alignment in gray matter, two tissues where the clear visual discriminability serves as ground truth. Secondly, our results indicate that our technique is even able to capture subtle variations of degree of alignment and fiber orientation within the white matter.



\section{Materials and methods}
\label{sec:section2}

\begin{figure}[htbp]
	\centering
	\includegraphics[scale=0.4]{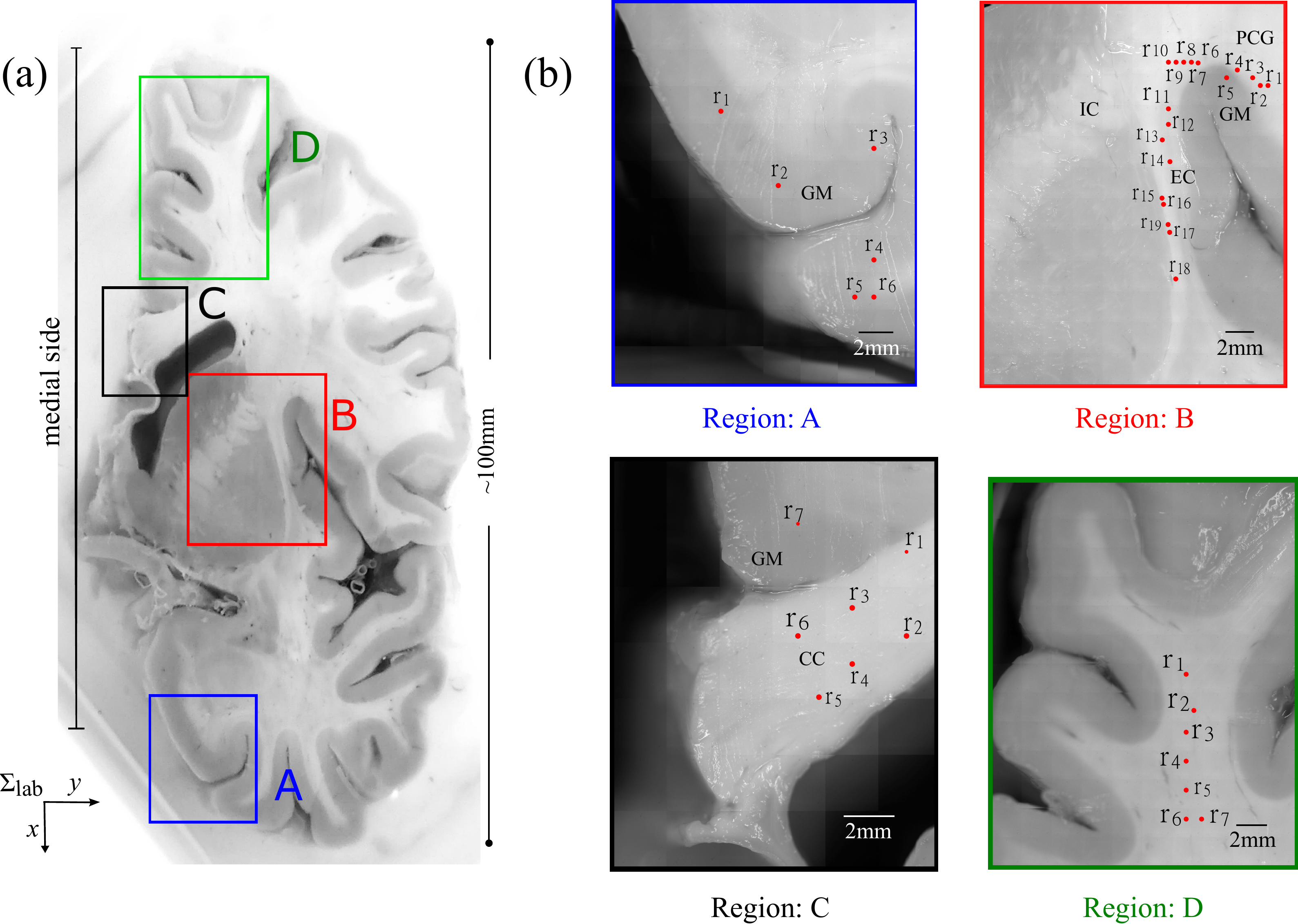}
	\caption{\textbf{(a)} Photograph of a coronal cross section of a healthy human brain fixed in $4\%$ formaldehyde solution. The regions marked in different colors were probed in this study. The medial side of the brain was aligned parallel to the x-axis in the output coordinate system ($\sum_{out}$). \textbf{(b)} shows  zoomed in images of the regions marked as A, B, C and D in (a). The red points illustrate the illumination spots of the probing beam (measurement points $\textbf{r}_{i}$). GM: Gray Matter, CC: Corpus Callosum, EC: External Capsule, IC: Internal Capsule and PCG: Pre-Central Gyrus. }
	\label{fig:brain}
\end{figure}

\subsection{Brain sample} 
\label{sec:section2a}

Two types of tissues can visually easily be distinguished within the brain, i.e., gray matter and white matter (see Fig.~\ref{fig:brain}(a)). The gray matter, also called cortex, constitutes the surface of the brain and is home to the cell bodies of the neurons. Multiple folds, called gyri, increase the surface of the brain. The white matter below the gray matter is made up of nerve fibers  (axons) through which different neurons interact with each  other. Nerve fibers that reach neurons in neighboring gyri are called U-fibers. Nerve fibers that reach neurons further away on the same hemisphere of the brain are called association fibers and those  connecting the two  hemispheres of the brain are called commissural fibers. Finally, projection fibers link the white matter to the spinal cord. For better electric conductivity along the fibers, most nerve fibers are isolated through sheaths of myelin, which are responsible for the white color of the white matter and for the birefringent properties of the nerve fibers ~\cite{Larsen2007}. 

Fig.~\ref{fig:brain}(a) shows a coronal section of a healthy human brain (anonymous donor). The sample was obtained during autopsy and was fixed in neutral buffer formalin ($4\%$ formeldehyde) solution. For our investigation, we have selected distinct areas within this section: A, B, C and D.   The corpus callosum (region C) consists of commissural fibers that connect the left and right hemisphere of the brain. The corpus callosum is the largest fiber tract system in the brain, along its axis all the fibers are aligned. The external capsule, a thin elongated structure (region B), consists of projection fibers connecting the frontal lobe and the spinal cord, as well as to a lesser degree of association fibers that interconnect the frontal and the parietal cortex. Inside the external capsule, a distinct orientation of fibers along the axis of this elongated structure is thus found. Whereas the corpus callosum and the external capsule each represent a well-aligned fiber system, the white matter inside the gyri contains connections to various parts of the brain and thus a mixture of U-fibers, association,  commissurial and projection fibers. This results in fibers aligned at different depths in different directions and often over several levels with a mean fiber orientation  towards the center of the brain~\cite{Oishi2011}. The superior frontal gyrus (region D) was chosen to showcase this situation. Whereas region B to D were chosen to allow the investigation of distinctive features of fiber alignment in white matter, we additionally selected  region A to investigate the response of gray matter.

\begin{figure}[htbp]
	\centering
	\includegraphics[scale=0.5]{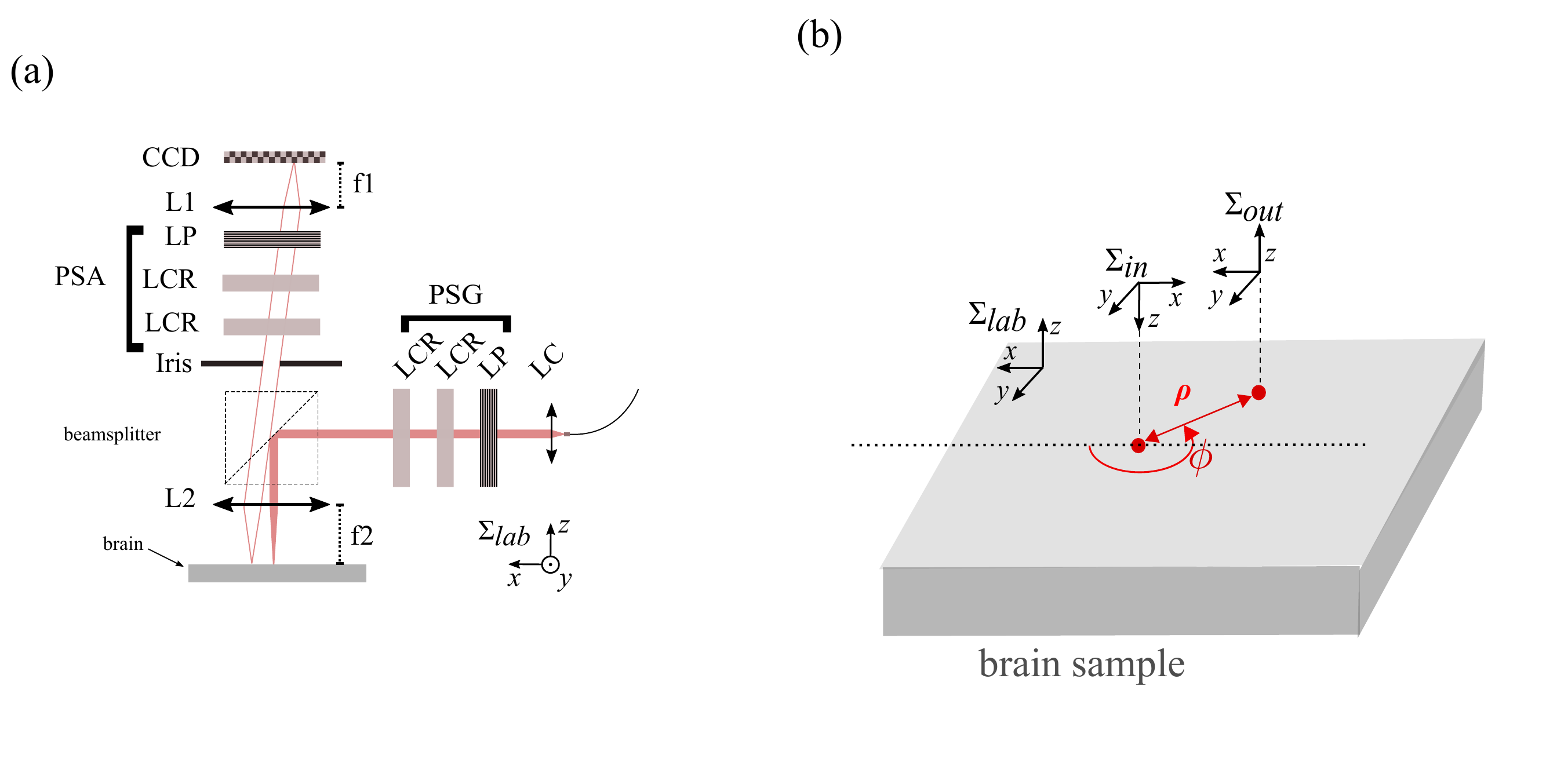}
	\caption{\textbf{(a)}  Schematic drawing (not to scale) of the polarimetric imaging setup used to
		probe the brain sample in the backscattering geometry. This instrument has
		two arms: an illumination and a detection arm containing polarization state generator
		(PSG) and polarization state analyzer (PSA), respectively. LC, L1, and L2 denote lenses. LCR are liquid crystal retarders and LP are linear polarizers. The tissue sample was probed at $\lambda = 700\,nm$ with a focused beam. The backscattered intensity distribution
		is recorded by a CCD camera. $\sum_{lab}$ denotes the
		laboratory frame. \textbf{(b)}  Schematic representation of the orientation of the right-handed coordinate systems involved in the experimental setup, relative to the brain sample. The input polarization state is defined in the input coordinate system ($\sum_{in}$). The detection frame $\sum_{out}$ attached to the CCD chip coincides with $\sum_{lab}$. For the radial analysis carried out in this study, each	point on the probed surface is defined by the polar coordinates $(\rho, \phi)$, where the probing beam was focussed at $\textbf{r}(\rho = 0, \phi = 0)$.}
	\label{fig:brain2}
\end{figure}

\subsection{The polarimetric microscope}
\label{sec:section2b}

The polarimeteric microscope used to probe the coronal section of the healthy human brain is described in detail~\cite{Hornung2019,Jain2020} and was slightly adapted for this study. An objective lens (L2) of focal length $f=30$\,mm (NA = 0.13) was used to focus the polarized illumination beam onto the tissue surface. The beam's spot radius ($1/e^2$) was measured to be $43$\,$\mu$m. The backscattered light was guided through the two-lens system composed of the objective lens L2 and a second lens L1 (focal length $f=100$\,mm).  
A CCD chip  (ptGray grasshopper, 16bit, $2448\times2048$ pixels) positioned at the end of the detection arm was used to image (with a magnification of $\sim 3.1$) the spatial distribution of the backscattered light from the sample surface (field of view 500 x 500 $ \mu $$ m^2$).  
All measurements presented were performed at $\lambda=$700\,nm. A computerized scanning stage  (H101P2BX ProScan stage operated via a V31XYZE controller, both from PRIOR Scientific) was used to scan the sample in $x,y-$direction.

\subsection{Measurement procedure}
\label{sec:section2c}

The tissue sample was taken out from the $4\%$ formaldehyde solution and placed on a Petri dish. It was then positioned under the polarimetric microscope such that the medial part of the brain was aligned parallel to the $x-$axis in the lab coordinate system ($\sum_{lab}$), as depicted in Fig.~\ref{fig:brain}. To ascertain a consistent spatial registration of different measurement points, all the measurements were performed with the same placement of the sample.
The backscattered intensity distribution, recorded for different pairs of illumination and detection states, was averaged 10 times for noise reduction. For each measurement point, speckle noise was further reduced by repeating the measurement at four different beam positions located on the corners of a $5\,\mu m$ square region and again averaging the results. In addition to this, the background noise/dark current of the CCD was subtracted. For each region (A to D) on the brain sample, we probed various points $\textbf{r}_{i}$. For visualization of the position of the measurement points relative to the anatomy of the brain sample, we overlay the points onto the widefield images of the tissue surface taken using white light illumination from the polarimetric microscope.

\subsection{Analysis of the polarimetric data}
\label{sec:subsection2d}
The methodology applied in this study is discussed in detail in~\cite{Hornung2019,Jain2020}.  In~\cite{Jain2020}, we analyzed the polarimetric response from fibrous membranes (electrospun PVDFhfp scaffold) with varying degree of alignment to understand how the structural anisotropy manifests itself in the polarimetric response. More precisely, we analyzed the backscattered Stokes vector $\vec{S}$, from which we calculated the four polarization ellipse parameters (PEPs; orientation, ellipticity, helicity and degree of polarization). This results in spatially resolved maps of PEPs, which are conveniently defined in a polar coordinate system $(\rho, \phi)$, where $\rho$ is the distance from the probing beam center, and $\phi$ is the angle relative to the x-axis of the output coordinate system. Among the various PEPs, the spatial distribution of the degree of linear polarization (DOLP) $\Pi_{L}(\rho,\phi)$ gives a most intuitive understanding of the fibrous samples' degree of alignment. To simplify the analysis, we performed a radial averaging of the $\Pi_{L}(\rho,\phi)$-image, resulting in $\left\langle \Pi_L(\phi) \right\rangle_{\rho}$. 
By analyzing the dependence of $\phi_{max}$ with respect to the input stokes vector $\vec{S}_{in}$, a correlation between anisotropy and birefringence can be established. $\phi_{max}$ is defined as the direction $\phi$ for which maximum linear states are found in $\left\langle \Pi_L(\phi) \right\rangle_{\rho}$. In addition, the dependence of the peak value of $\left\langle \Pi_L(\phi=\phi_{max}) \right\rangle_{\rho}$ with respect to the orientation of $\vec{S}_{in}$ is also analyzed.
To determine $\phi_{max}$ and $\left\langle \Pi_L(\phi_{max}) \right\rangle_{\rho}$, we fit a Gaussian model.
To estimate the mean alignment direction of the fibers in the membranes, we analyze the orientation $\psi(\rho,\phi)$ parameter in the PEPs~\cite{Jain2020}, i.e. the orientation of the major axis of the polarization ellipse.

\section{Results and discussion}
\label{sec:section3}

\begin{figure}[htbp]
	\centering
	\includegraphics[scale=0.45]{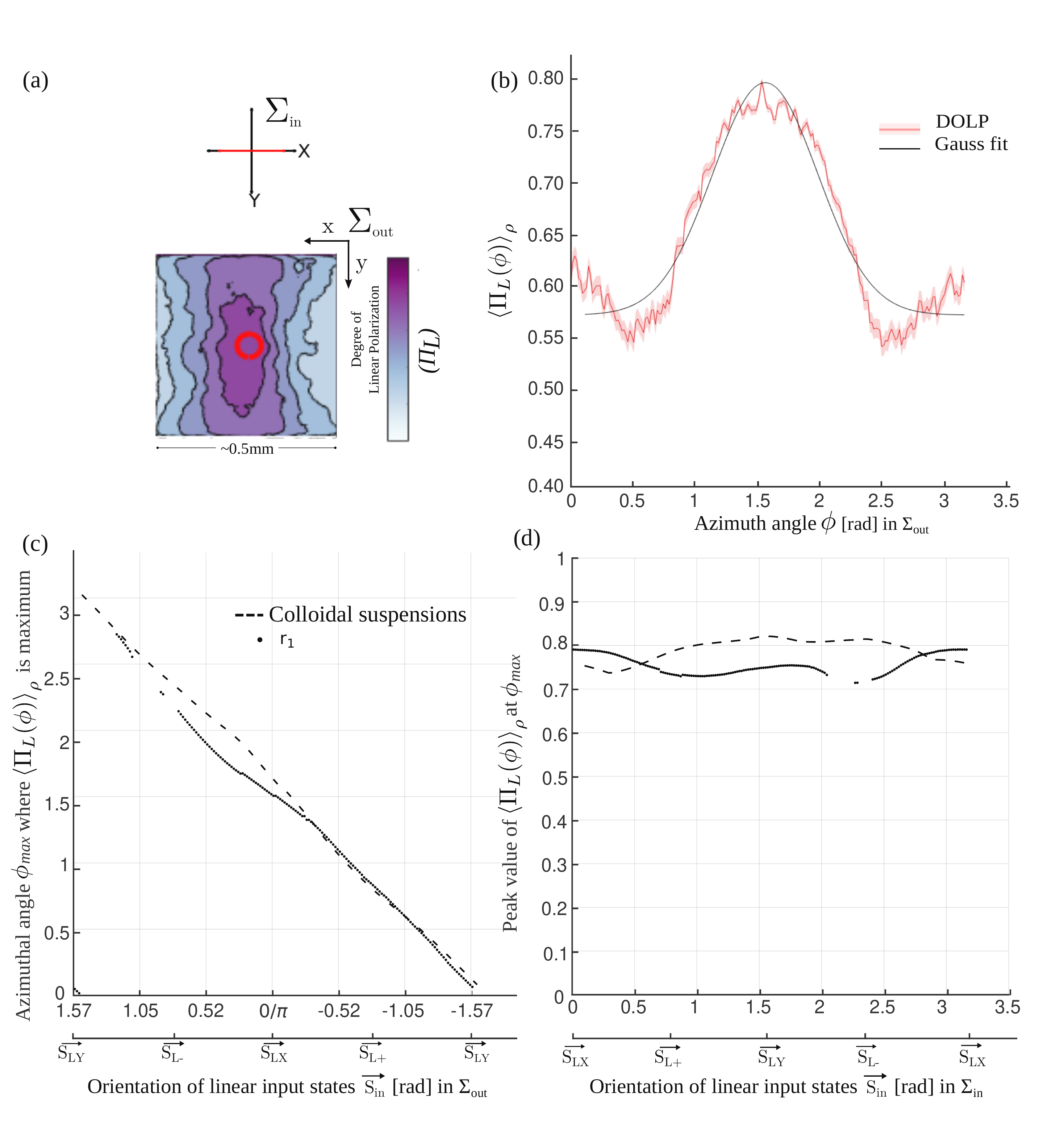}
	\caption{Exemplary result for gray matter, point $\textbf{r}_{1}$ in region A. \textbf{(a)} The spatial distribution of the $\Pi_{L}(\rho,\phi)$- image for the input polarization state depicted in red (in $S_{in}=x$), is shown in the output coordinate system ($\sum_{out}$). The red circle at the center of the image represents the focal spot size and position of the probing beam. \textbf{(b)} The radial analysis of the  $\Pi_{L}(\rho,\phi)$- image, for the input  state in red (in $\sum_{in}$, $S_{in}=0$). Video 1 (supplementary work) shows the $\Pi_{L}(\rho,\phi)$- image and video 2 (supplementary work)  illustrates the radial analysis for different input polarization state ($S_{in}=0\, to\, \pi$). \textbf{(c)} Azimuthal angle $\phi_{max}$ at which the maximum of $\left\langle \Pi_L(\phi) \right\rangle_{\rho}$ was recorded for different orientations (between $\pi/2$ and $-\pi/2$ in  $\Sigma_{out}$) of the linear polarization $\vec{S}_{in}$ of the probing beam. \textbf{(d)} Peak value of $\left\langle \Pi_L(\phi) \right\rangle_{\rho}$ recorded at $\phi_{max}$ for different orientations (between $0$ and $\pi$ in $\Sigma_{in}$) of the linear polarization $\vec{S}_{in}$ of the probing beam  ($\vec{S}_{in}$ is given in different coordinate systems for better representation of the outcomes).
	$\phi_{max}$ and $\left\langle \Pi_L(\phi=\phi_{max}) \right\rangle_{\rho}$ values are left blank for states $\vec{S}_{in}$ where no peak was identifiable in the radial analysis of the $\Pi_L(\rho,\phi)$-images. For reference, we performed the same radial analysis on measurements collected from colloidal suspensions~\cite{Hornung2019} (dashed black lines), to showcase quasi-perfect rotational symmetry.}
	\label{fig:300719_profile_process}
\end{figure}

\begin{figure}[htbp]
	\centering
	\includegraphics[scale=0.45]{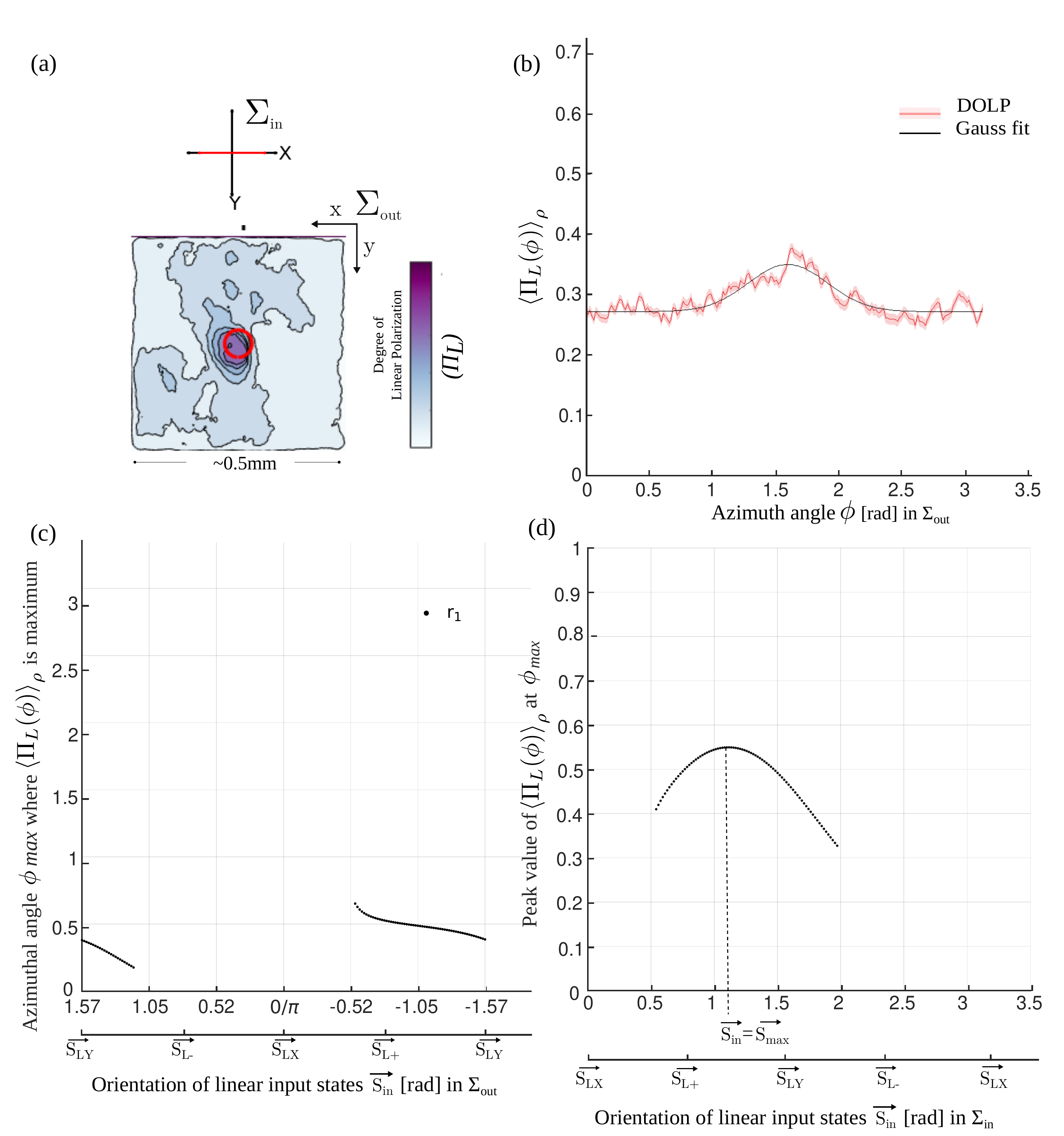}
	\caption{Exemplary result for white matter, point $\textbf{r}_{1}$ in region C. \textbf{(a)} The spatial distribution of the $\Pi_{L}(\rho,\phi)$- image for input polarization state depicted in red (in $\sum_{in}$, $S_{in}=0$), is shown in the output coordinate system $\sum_{out}$. The red circle at the center of the image represents the focal spot size and position of the probing beam. \textbf{(b)} The radial analysis of the  $\Pi_{L}(\rho,\phi)$- image, for the input  state in red. Video 3 (supplementary work) shows the $\Pi_{L}(\rho,\phi)$- image and video 4 (supplementary work) illustrates the radial analysis for different input polarization state ($S_{in}=0\, to\, \pi$). \textbf{(c)} Azimuthal angle $\phi_{max}$ (dotted line)at which the maximum of $\left\langle \Pi_L(\phi) \right\rangle_{\rho}$ was recorded for different orientations (between $\pi/2$ and $-\pi/2$ in  $\Sigma_{out}$) of the linear polarization $\vec{S}_{in}$ of the probing beam. \textbf{(d)} Peak value of $\left\langle \Pi_L(\phi) \right\rangle_{\rho}$ (dotted line)recorded at $\phi_{max}$ for different orientations (between $0$ and $\pi$ in $\Sigma_{in}$) of the linear polarization $\vec{S}_{in}$ of the probing beam ($\vec{S}_{in}$ is given in different coordinate systems for better representation of the outcomes).
			$\phi_{max}$ and $\left\langle \Pi_L(\phi=\phi_{max}) \right\rangle_{\rho}$ values are left blank for states $\vec{S}_{in}$ where no peak was identifiable in the radial analysis of the $\Pi_L(\rho,\phi)$-images.  For anisotropic sample the peak depicting the preservation of maximum linear states is $\vec{S}_{in}=\vec{S}_{max}$.}
	\label{fig:150719_profile_process}
\end{figure}

\subsection{Polarimetric response for gray and white matter}
\label{sec:section3a}

The key difference between gray and white matter of the brain is that gray matter majoritarily comprises of nucleon cells while white matter comprises of nerve fibers. First we analyzed the polarimetric response of two exemplarly points representing gray (point $\textbf{r}_1$, region A) and white matter (point $\textbf{r}_1$, region C), respectively. 

Fig.~\ref{fig:300719_profile_process} shows the results for point $\textbf{r}_1$, region A (gray matter). In Fig.~\ref{fig:300719_profile_process}(a) we see the $\Pi_{L}(\rho,\phi)$- image recorded for the input polarization state in red (in input coordinate system $\sum_{in}$), and Fig.~\ref{fig:300719_profile_process}(b) shows its respective radial analysis and Gaussian fit. The response is isotropic, \textit{i.e.}, observed as symmetry with respect to rotation of the input polarization, termed "rotational symmetry" in the remainder of this study. 
This rotational symmetry becomes evident in the animation depicting the  spatial distribution of the $\Pi_{L}(\rho,\phi)$- image for different polarization of the probing beam ($\sum_{in}$) (see video 1, supplementary work). 
The symmetry observed in  video 1 is quantified in Fig.~\ref{fig:300719_profile_process}(c) and (d), where the azimuthal direction $\phi_{max}$ along which the pattern is stretched depends on the polarization state of the probing beam $\vec{S}_{in}$. The strength of linear states depicts a quasi-flat response with respect to $\vec{S}_{in}$ which means that the depolarization does not depend on the orientation of the imput state. The values are left blank for polarization probing states $\vec{S}_{in}$ where no peak was identifiable. For comparison, the response from a colloidal suspension (175\,nm radius polystyrene spheres diluted in water, see~\cite{Hornung2019}) showing a perfect rotational symmetry  has been plotted as dashed black lines in  Fig.~\ref{fig:300719_profile_process}(c) and (d).

Fig.~\ref{fig:150719_profile_process} shows the analogue results for white matter (point $\textbf{r}_1$, region C). 
In contrast to gray matter no rotational symmetry can be observed, which indicates anisotropy (see video 3, supplementary work). Fig.~\ref{fig:150719_profile_process}(c) provides a quantitative visualization of the anisotropic nature of the tissue, \textit{i.e.}, the azimuthal direction $\phi_{max}$ along which the patterns in the $\Pi_{L}(\rho,\phi)$-images are stretched does not significantly depend on the orientation of the probing beam’s polarization. Furthermore, the linearity of the probing beam’s polarization state $\vec{S}_{in}$ is preserved predominantly for input states with orientations of $\vec{S}_{in}$ that are parallel and perpendicular to $\phi_{max}$. 
In ~\cite{Jain2020} we observed the same co-occurrence of rotational anisotropy and preservation of linear states in aligned PVDFhfp electrospun fibrous membranes and argued that this behaviour reflects the uniaxial birefringence of the material. For the brain sample, we observe a distinctive peak around a specific input state $\vec{S}_{in}$, defined as $\vec{S}_{max}$, where maximum linear polarization states are preserved (Fig\,\ref{fig:150719_profile_process}(d)). This allows us to define the mean orientation of the nerve fibers.

\begin{figure*}[htbp]
	\centering
	\includegraphics[scale=0.7]{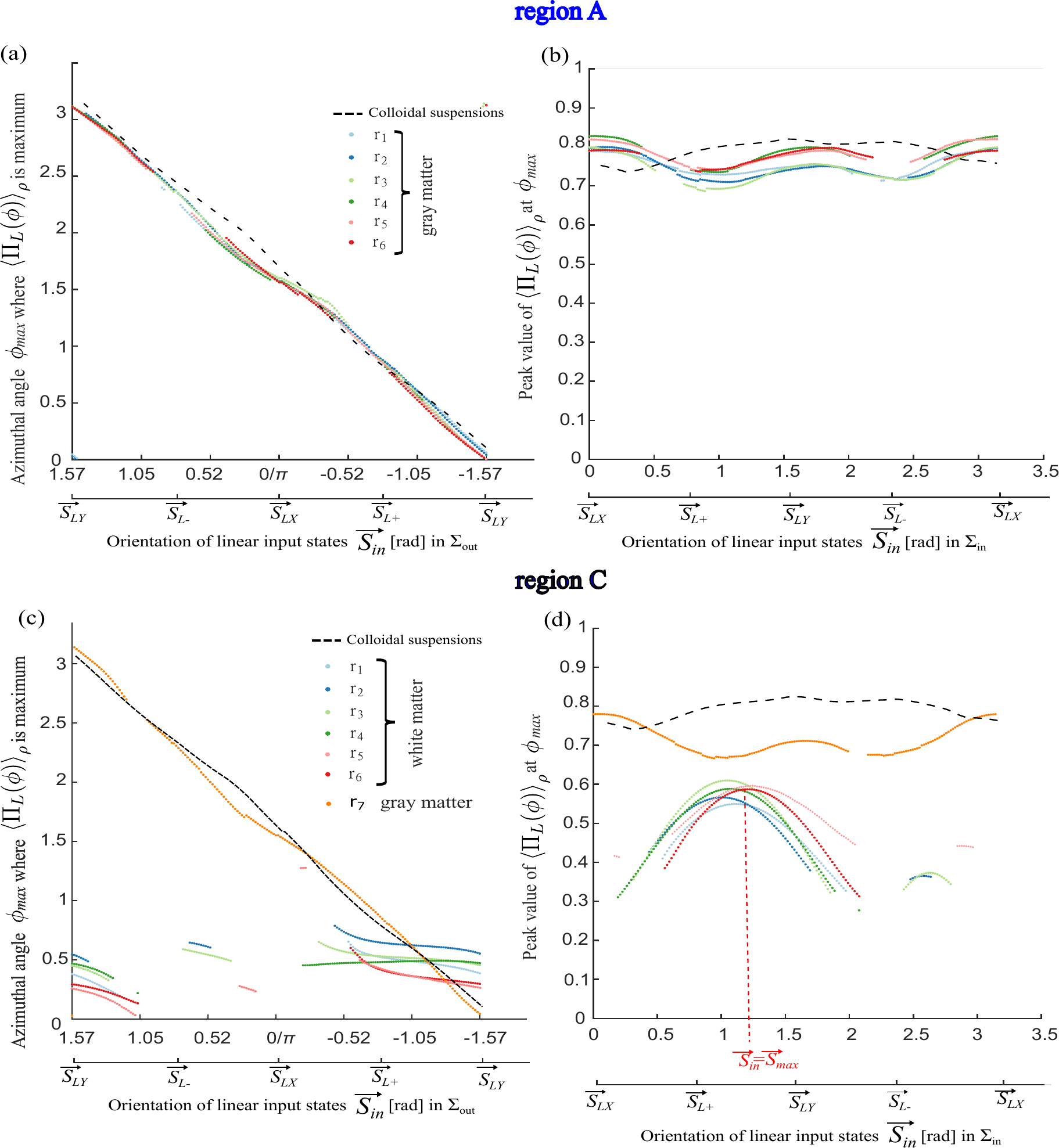}
	
	\caption{Summary of results from all points in region A and region C. \textbf{(a,c)} azimuthal angle $\phi_{max}$ at which the maximum of $\left\langle \Pi_L(\phi) \right\rangle_{\rho}$ was recorded for different orientations (between $\pi/2$ and $-\pi/2$ in $\Sigma_{out}$) of the linear polarization $\vec{S}_{in}$ of the probing beam.
			\textbf{(b,d)} peak value of $\left\langle \Pi_L(\phi) \right\rangle_{\rho}$ recorded at $\phi_{max}$ for different orientations (between $0$ and $\pi$ in $\Sigma_{in}$) of the linear polarization $\vec{S}_{in}$ of the probing beam ($\vec{S}_{in}$ is given in different coordinate systems for better representation of the outcomes).
			$\phi_{max}$ and $\left\langle \Pi_L(\phi=\phi_{max}) \right\rangle_{\rho}$ values are left blank for states $\vec{S}_{in}$ where no peak was identifiable in the radial analysis of the $\Pi_L(\rho,\phi)$-images. For anisotropic samples, $\vec{S}_{in}=\vec{S}_{max}$ is the input polarization state for which maximum linear output states are obtained (see \textbf{(d)}.
			As a reference of rotational symmetry, the same radial analysis is shown for a colloidal suspension~\cite{Hornung2019} (dashed black lines).
	}
	\label{fig:300719_profiles}
\end{figure*}

Fig.~\ref{fig:300719_profiles} shows the responses for various different measurement points ($\textbf{r}_{i}$) in regions A and C, respectively. These results showcase the reproducibility of rotational symmetry in gray matter (\textit{i.e.} all the points in region A and point $r_{7}$ in region C), and anisotropy in white matter (\textit{i.e.} for points $\textbf{r}_{i}$, for i=1...6 in region C). 

\begin{figure}[htbp]
	\centering
	\includegraphics[scale=0.7]{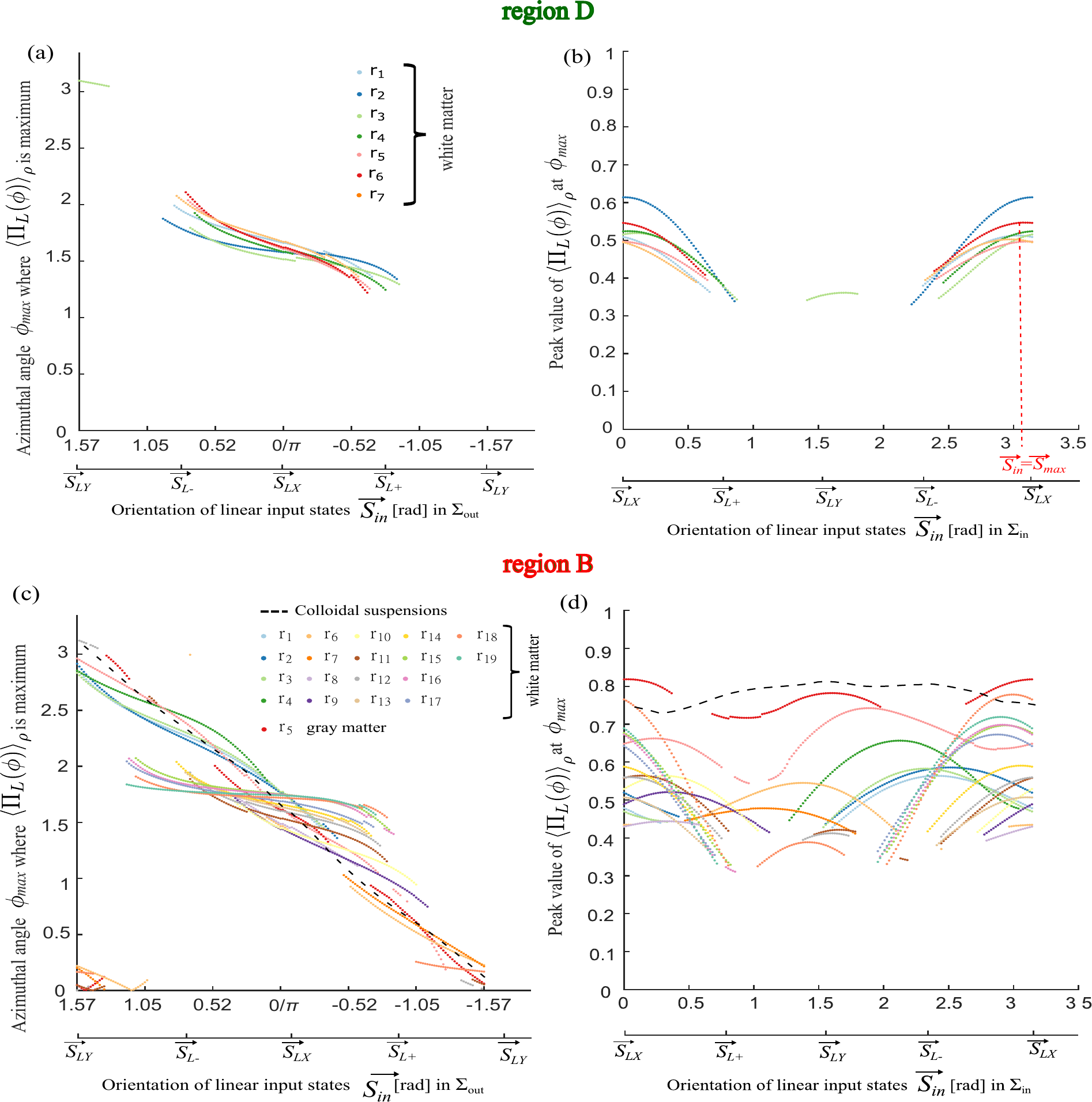}
	\caption{Summary of results from all points in region D and region B. \textbf{(a,c)} azimuthal angle $\phi_{max}$ at which the maximum of $\left\langle \Pi_L(\phi) \right\rangle_{\rho}$ was recorded for different orientations (between $\pi/2$ and $-\pi/2$ in $\Sigma_{out}$) of the linear polarization $\vec{S}_{in}$ of the probing beam .
		\textbf{(b,d)} peak value of $\left\langle \Pi_L(\phi) \right\rangle_{\rho}$ recorded at $\phi_{max}$ for different orientations (between $0$ and $\pi$ in $\Sigma_{in}$) of the linear polarization $\vec{S}_{in}$ of the probing beam ($\vec{S}_{in}$ is given in different coordinate systems for better representation of the outcomes).
		$\phi_{max}$ and $\left\langle \Pi_L(\phi=\phi_{max}) \right\rangle_{\rho}$ values are left blank for states $\vec{S}_{in}$ where no peak was identifiable in the radial analysis of the $\Pi_L(\rho,\phi)$-images.
		For anisotropic samples, $\vec{S}_{in}=\vec{S}_{max}$ is the input polarization state for which maximum linear output states are obtained. As a reference of rotational symmetry, the same radial analysis is shown for a colloidal suspension~\cite{Hornung2019} (dashed black lines).
		}
		
	\label{fig:160719_profiles}
\end{figure}

Similarly, for regions D and B, the resulting curves for different measurement points are presented in Fig.~\ref{fig:160719_profiles}.

In all points in all areas, the results are akin to those of the electrospun PVDFhfp scaffolds and colloidal suspensions~\cite{Jain2020}. In the gray matter, where a low degree of alignment is expected due to the quasi-random intermixing of cell bodies, dendrites, and nerve fibers, the polarimetric response is that of an isotropically scattering medium. In the white matter, where a significant degree of alignment is expected due to the alignment of nerve fibers, the polarimetric response resembles that of a uniaxially birefringent medium.

\subsection{Estimating the mean alignment direction and degree of alignment of nerve fibers}
\label{sec:section3b}

In our previous study on electrospun PVDFhfp scaffolds~\cite{Jain2020} we could show that in samples that exhibit a certain degree of alignment linear states are maximally preserved for input polarization $\vec{S}_{in}$ parallel and perpendicular to the mean fiber alignment direction, and that the direction $\phi_{max}$ (along which the $\Pi_L(\rho,\phi)$-image is stretched) is perpendicular to the fiber orientation. In the present study, where the mean alignment direction of nerve fibers in the white matter is a priori unknown, we take the direction of the probing state $\vec{S}_{max}$ (which is roughly perpendicular to $\phi_{max}$) as a proxy for the mean alignment direction. Then, in order to refine this quantification, we analyze the orientation parameter $\psi(\rho,\phi)$, i.e. the PEP which describes the orientation of the major axis of the polarization ellipse. The mean alignment direction is estimated as the mean orientation parameter over  $\rho$ along  $\phi=\phi_{max}$. 

The curves in Fig.~\ref{fig:300719_profiles}(a,c) and Fig.~\ref{fig:160719_profiles}(a,c) provide a quantitative indicator for the degree of alignment, but their interpretation so far relies on visual inspection of the slope of the curves. 
Thereby, a slope near -1 indicates a low degree of alignment, while a slope near 0 indicates a high degree of aligment~\cite{Jain2020}. 
For a quantitative assessment of the slope we calculated the derivative at every input state and took the median of its distribution, which we define as the slope parameter $\mathscr{m}$. In Table\,\ref{table:1} the values of  $\mathscr{m}$ are given for each $\textbf{r}_{i}$ in the regions A-D. 

\begin{figure}[htbp]
	\centering
	\includegraphics[scale=0.38]{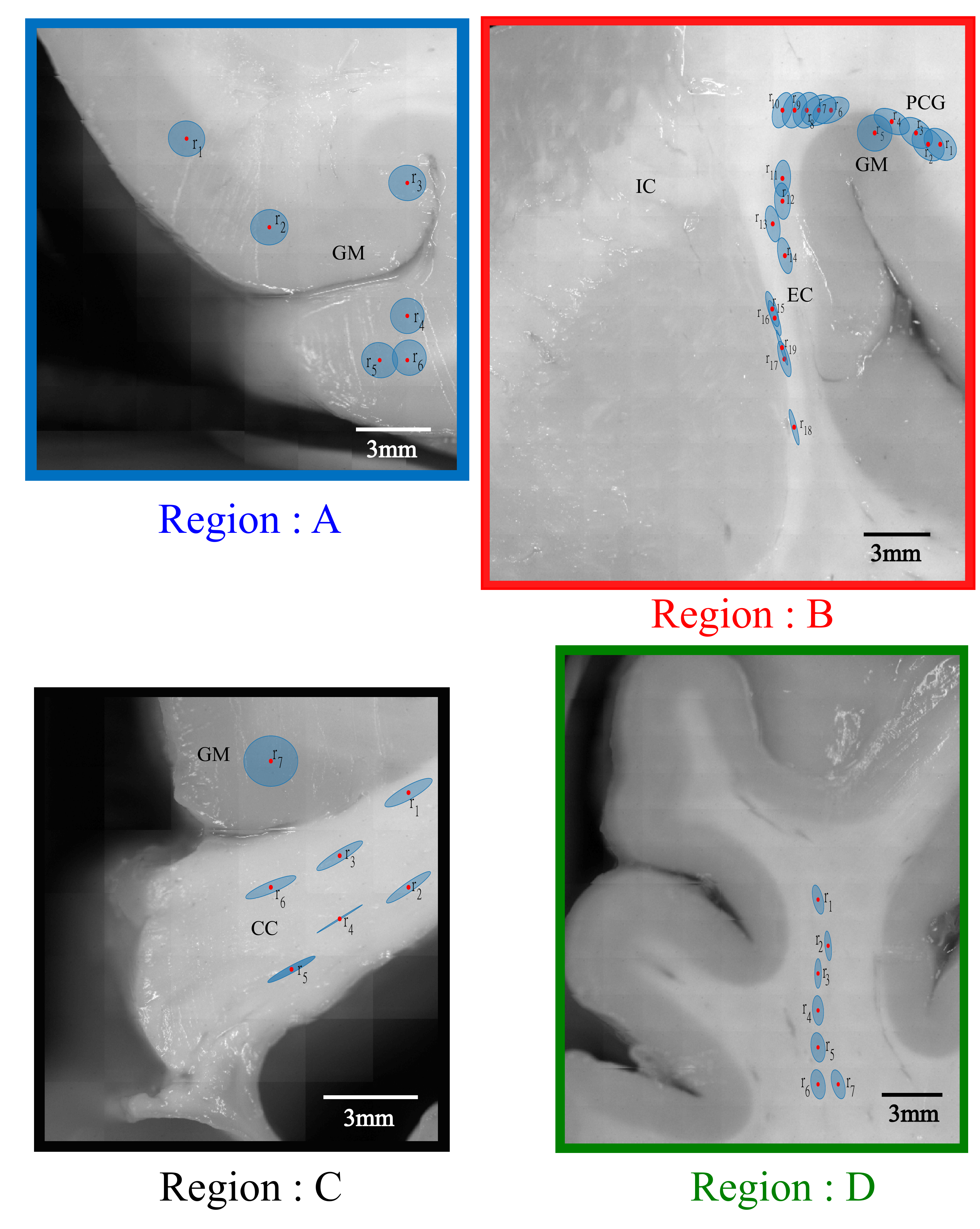}
	\caption{Ellipse-encoded visualization of mean alignment direction and degree of alignment for different measurement points ($\textbf{r}_{i}$) in regions (A-D).  The ellipses are superimposed on the white light widefield image of the tissue surface. The red circles represent the position of the probing beam. The orientation of the ellipses encodes the estimated mean alignment direction and the aspect ratio encodes the degree of alignment. For high degree of alignment the ellipse approaches a line, while for zero degree of alignment it becomes a circle (\textit{eg.} region: A). GM: Gray Matter, CC: Corpus Callosum, EC: External Capsule, IC: Internal Capsule and PCG: Pre-Central Gyrus. Note that the length of the ellipse axis can vary from one region to another due to the different scalings.}
	\label{fig:brain_orientations}
\end{figure}

To allow a direct visual comparison with the anatomical context, the values of $\mathscr{m}$ together with the mean orientations are coded in a graphical representation, i.e. as ellipses, and overplotted onto the widefield image in Fig. \ref{fig:brain_orientations}. The orientation of the major axis in the ellipse denotes the mean alignment direction, while the size of the minor-axis relative to the major axis depicts the value of $\mathscr{m}$. Thereby, the minor-axis of the ellipse was defined as the length of the major-axis (length of the major axis is a constant arbitrary value chosen for graphical representation) multiplied by the absolute value of $\mathscr{m}$.

\begin{table}[htbp]
	\centering
	\begin{tabular}{ccccc}
		\multicolumn{5}{c}{Slope parameter $(\mathscr{m})$}                                                                                                                                                                                                                                                               \\
		\multicolumn{1}{c|}{\textbf{\textcolor{blue}{A}}} & \multicolumn{2}{c|}{\textbf{\textcolor{red}{B}}}   & \multicolumn{1}{c|}{\textbf{C}} & \textbf{\textcolor[rgb]{0,0.502,0}{D}} \\ \hline
		\multicolumn{1}{c|} {\textbf{$\textbf{r}_{1}$ : -0.99}}                                                & \multicolumn{1}{c|}{$r_{1}$ : -0.69}                                               &
		\multicolumn{1}{c|}{$r_{10}$ : -0.48} & \multicolumn{1}{c|}{$r_{1}$ : -0.25}           & $r_{1}$ : -0.37 \\
		\multicolumn{1}{c|}{\textbf{$\textbf{r}_{2}$ : -1.02}}                                                & \multicolumn{1}{c|}{$r_{2}$ : -0.72}                                               &
		\multicolumn{1}{c|}{$r_{11}$ : -0.42} & \multicolumn{1}{c|}{$r_{2}$ : -0.18}           & $r_{2}$ : -0.24                                                              \\
		\multicolumn{1}{c|}{\textbf{$\textbf{r}_{3}$ : -1.00}}                                                & \multicolumn{1}{c|}{$r_{3}$ : -0.67}                                               &
		\multicolumn{1}{c|}{$r_{12}$ : -0.41}  & \multicolumn{1}{c|}{$r_{3}$ : -0.20}           & $r_{3}$ : -0.25 \\
		\multicolumn{1}{c|}{\textbf{$\textbf{r}_{4}$ : -0.89}}                                                & \multicolumn{1}{c|}{$r_{4}$ : -0.61}                                               &
		\multicolumn{1}{c|}{$r_{13}$ : -0.36}  & \multicolumn{1}{c|}{$r_{4}$ : -0.02}           & $r_{4}$ : -0.41                                                              \\
		\multicolumn{1}{c|}{\textbf{$\textbf{r}_{5}$ : -0.96}}                                                & \multicolumn{1}{c|}{\textbf{$\textbf{r}_{5}$ : -0.89}}                                               &
		
		\multicolumn{1}{c|}{$r_{14}$ : -0.34}   & \multicolumn{1}{c|}{$r_{5}$ : -0.27}           & $r_{5}$ : -0.52                                                              \\
		\multicolumn{1}{c|}{\textbf{$\textbf{r}_{6}$ : -0.95}}                                                & \multicolumn{1}{c|}{$r_{6}$ : -0.63}                                               &
		\multicolumn{1}{c|}{$r_{15}$ : -0.22}   & \multicolumn{1}{c|}{$r_{6}$ : -0.21}           & $r_{6}$ : -0.52                                                              \\
		\multicolumn{1}{c|}{}                                                                  & \multicolumn{1}{c|}{$r_{7}$ : -0.73}                                               &
		\multicolumn{1}{c|}{$r_{16}$ : -0.19}    & \multicolumn{1}{c|}{\textbf{$\textbf{r}_{7}$ = -0.98}}           & $r_{7}$ : -0.46                                                              \\
		\multicolumn{1}{c|}{}                                                                  & \multicolumn{1}{c|}{$r_{8}$ : -0.68}                                               &  \multicolumn{1}{c|}{$r_{17}$ : -0.25} &                                                &                                                                                \\
		\multicolumn{1}{c|}{}                                                                  & \multicolumn{1}{c|}{$r_{9}$ : -0.60}                                               & \multicolumn{1}{c|}{$r_{18}$ : -0.13}  &                                                  &                            \\
		&   \multicolumn{1}{c|}{}        & \multicolumn{1}{c|}{$r_{19}$ : -0.08}                                              &                                                  &                                                         
		
	\end{tabular}
	\caption{Slope parameter $(\mathscr{m})$, quantified from the curves in Fig.~\ref{fig:300719_profiles}(a,c) and \ref{fig:160719_profiles}(a,c) of each measurement point: $\textbf{r}_{i}$ in Fig.~\ref{fig:brain}. The value of $\mathscr{m}$ varies from 0 (indicating a high degree of alignment) to -1 (indicating a zero degree of alignment). Bold = gray matter, regular = white matter}
	\label{table:1}
\end{table}		

In all points $\textbf{r}_{i}$ located inside the gray matter, our results indicate a very low to zero degree of alignment. This is visualized by the circularity of the ellipses, which corresponds to  $\mathscr{m}$ values persistently found around  $\mathscr{m}$ = -1, i.e. in-between -0.89 and -1.02 (see table \ref{table:1}, bold values).
In comparison, for points located in white matter the ellipses deviate substantially from the circular shape ($\mathscr{m}$ values range from -0.03 to -0.73). 
In the most extreme case, inside the corpus callosum in Fig.\,\ref{fig:brain_orientations} region C, the ellipses are  flat close to a line, representing low and fairly constant $\mathscr{m}$ values between -0.03 and -0.27. Such strong degree of orientation can be expected since the corpus callosum contains a rather uniform density of well aligned in-plane nerve fibers running parallel to a common axis ~\cite{Oishi2011, Menzel2020}. 

The measurements within the white matter in region D provide a more differentiated picture reflecting its anatomical perculiarities. The fine local differences in $\mathscr{m}$ showcase the sensitivity of the $\mathscr{m}$ values to small variations in tissue architecture.
The mean orientation indicated by the main axis of the ellipses runs parallel to a virtual trajectory that reflects the direction of the nerve fibers running from the gyrus towards the center of the brain~\cite{Oishi2011}. Along this trajectory, the $\mathscr{m}$ values indicate a continuous variation of the degree of alignment. Starting from a medium degree of alignment near the bifurcation of white matter at ${r}_{1}$, the degree of alignment increases to a maximum at the narrowest part medial to this bifurcation (${r}_{2}$ and ${r}_{3}$), and then decreases again towards the next bifurcation (${r}_{4}$ to ${r}_{7}$). This behaviour can on one hand be explained by association, commissurial, and projection fibers that intersect with a larger variety of angles leading  to a lower degree of alignment at bifurcations than at the narrow part between sulci, but on the other hand by a higher prevalence of crossing U fibers oriented at different planes in areas where the white matter is thicker.

A similar observation of gradually changing orientation and ellipticity is made in region B. The degree of alignment is indicated to be strongest inside the external capsule (${r}_{15}$ to ${r}_{19}$, $\mathscr{m}$ values ranging between -0.25 and 0.08) with a mean orientation consistently pointing along the axis of this structure. Both observations agree well with the expected alignment of densely packed nerve fibers connecting the putamen and the claustrum. In comparison, the degree of alignment is indicated to be substantially lower around the sulcus towards the pre-central gyrus and the mean orientation in the different points follows the curvature of this sulcus (${r}_{1}$ to ${r}_{10}$, $\mathscr{m}$ values between -0.73 and -0.48). The transition is again continuous with a medium degree of alignment indicated in ${r}_{11}$ to ${r}_{14}$ located at the upper end of the external capsule. The low degree of alignment around the sulcus can be explained by a crossing of nerves fibers from the external capsule with the fibers of the corona radiate system exiting the internal capsule in the same area, and by fibers from the arcuate fasciculus running perpendicular to the section plane~\cite{Stieglitz2013}.

The observed qualitative correlation of $\mathscr{m}$ along fiber trajectories and their correspondence to anatomical structures in the brain indicate that our results are indeed sensitive to small variations in tissue architecture. A more quantitative assessment of the possible accuracy of this technique can be obtained when calculating the standard deviation (std) of $\mathscr{m}$ within groups of measurement points. For gray matter in region A, the std is 0.044, and for white matter in region C (corpus callosum), it is 0.088. In region B (external capsule), we distinguish four groups within which $\mathscr{m}$ is fairly constant. The std are 0.046 (${r}_{1}$ to ${r}_{4}$), 0.098 (${r}_{6}$ to ${r}_{10}$), 0.040 (${r}_{11}$ to ${r}_{14}$), and 0.070 (${r}_{15}$ to ${r}_{20}$). Assuming that the degree of alignment was constant within each of these groups, the std gives an upper bound of the achieved accuracy (whereas the actual accuracy can be even better, if part of the observed variations is explained by the tissue structure). Apart from two outliers, the std values are consistently found around 0.05, indicating that the measured differences between the different groups in region B are indeed significant (the mean values are 0.673, 0.642, 0.382, and 0.176) and correspond to actual variations of tissue architecture along the external capsule. 

\ 

\section{Conclusion}
\label{sec:section4}
In this study, we used polarimetric imaging to investigate the structural anisotropy of different regions in a coronal cross-section of a human brain. Our results demonstrate the possibility to clearly distinguish the two structurally distinct tissue types constituting the human brain, \textit{i.e.},  gray and white matter, by analyzing the tissue's backscattered polarimetric response. Here, the widefield images in which the two tissue types can visually be discerned serve as a reference.
By defining a slope parameter $\mathscr{m}$, which provides an intuitive understanding of the degree of alignment of the micro-architecture in the tissue, we observed that there is a strong correlation between the $\mathscr{m}$ values and the tissue structure. Values close to -1 are consistently found in gray matter, indicating isotropic behavior and rotational symmetry, which is expected due to the isotropic distribution of cell bodies in gray matter. Values mostly in the range between -0.5 and 0 are found in white matter, indicating a high degree of alignment, which corresponds to the prevalence of nerve fiber bundles. The estimated mean alignment direction found in white matter is in agreement with the nerve fibers pathways in the brain ~\cite{Oishi2011,Zilles2016}.
Further, the analysis of regions B and D suggests that we can  sensitively distinguish small nuances in the degree of alignment of nerve fibers in the white matter. This allows us to visualize the gradual transition of $\mathscr{m}$ from regions with high degree of alignment where associaction, commissural, and projection fibers run parallely in one direction, to regions of lower degree of alignment where the fibers are intertwined, located in different planes and where U fibers are present.  
The assumption that these qualitative observations relate to actual tissue architecture is supported by the analysis of the std of $\mathscr{m}$ within spatial groups. Assuming that the degree of alignment does not vary significantly within these areas, the std serves as an upper bound of the achievable accuracy of determining $\mathscr{m}$. 

In conclusion, the results suggest that the proposed technique  allows to sensitively define the different degrees of alignment in healthy brain tissue in-vivo without prior tissue preparation. On one hand, it could be used to help accurate delineation of cancer tissue during resection. On the other hand, given the promising sensitivity of distinguishing nuances in degree of alignment inside white matter, it could be used to guide resections by identifying vital nerve fiber trajectories by performing tractography of nerve fiber pathways in the brain. In addition to applications in the brain, the presented method has potential to identify pathologies in other tissue types consisting of fibrous structures, such as  the skin for diagnosing skin cancer or for monitoring the healing process after skin injuries.







\section*{Acknowledgment}

The Authors would like to thank René Nyffenegger and Manes Hornung for their valuable inputs and Ekkehard Hewer from the Institute of Pathology, University Hospital Bern for providing the brain sample. We thank Lynn Roth and André Stefanov for many stimulating discussions and proof-reading and Florentin Spadin for helping with the figures.


\bibliography{arushimanuscript}

\end{document}